\documentclass{aa}  

\usepackage{graphicx}
\usepackage{txfonts}
\usepackage{lipsum}
\usepackage{subcaption}         
\usepackage{lscape}             
\usepackage{placeins}           
                                
\begin{document}

   \title{Cyclic sunspot activity during the first millennium CE as reconstructed from radiocarbon}


%

   \author{I. Usoskin\inst{1,3}\fnmsep\thanks{Corresponding author: ilya.usoskin@oulu.fi}
        \and S.K. Solanki\inst{2}
        \and N.A. Krivova\inst{2}
        \and T. Chatzistergos\inst{2}
        }
\institute{Space Physics and Astronomy research Unit, University of Oulu, Finland 
   \and Max-Planck Institute for Solar System Research, Justus-von-Liebig-Weg 3, 37077 G\"ottingen, Germany
   \and Institute for Space-Earth Environmental Research, Nagoya University, Furo-cho, Chikusa-ku, Nagoya 464-8601, Japan}

   \date{Received September 30, 20XX}

 
  \abstract
   {Solar activity, dominated by the 11-year cyclic evolution, has been observed directly since 1610 CE. Before that, indirect cosmogenic proxy data are used to reconstruct it over millennia. Recently, the precision of radiocarbon $\Delta^{14}$C measurements has improved sufficiently to allow reconstructing cyclic solar activity over millennia.}
   {We present the first detailed reconstruction of solar activity, represented here by annual sunspot numbers, during the first millennium, 1\,--\,969, CE. 
   } 
   {The reconstruction of sunspot numbers from $\Delta^{14}$C was performed using a physics-based method which involves several steps: using the modern carbon-cycle box model, the $^{14}$C production rate, corrected for the contemporary geomagnetic shielding, was computed from the measured concentrations; The open solar magnetic flux was computed using a model of the heliospheric cosmic-ray modulation; Sunspot numbers were calculated by inverting a model of the evolution of the Sun's magnetic field. The Markov Chain Monte Carlo approach was used to directly account for different sources of uncertainty.}
   {Annual sunspot numbers were reconstructed for the ﬁrst millennium CE. 
   This period includes one extreme solar event occurring in 774 CE and one Grand solar minimum of 650\,--\,730 CE.
   We could identify 91 solar cycles, of which 26 were well-defined, while 24, and 41 were reasonably and poorly defined, respectively.
   The mean cycle length was 10.6 years, but the lengths of individual cycles vary between 8 and 15 years. 
   The existence of empirical Waldmeier's relations remains inconclusive with this dataset.
   No significant periodicities were found beyond the 11-year cycle.}
   {This work fills the gap in the solar cycle statistics between the previously reconstructed first millennium BCE and the second millennium CE, providing vital constraints for the solar dynamo and irradiance models. 
   A consistent 3-millennium-long reconstruction of sunspot numbers, based on a composite multi-proxy cosmogenic record, is pending.}


   \maketitle
\nolinenumbers

\section{Introduction}
\label{S:Intro}

The Sun is a variable star, whose variability is an important driver of the terrestrial environment \citep[e.g.,][]{gray10,haigh10}, but also serves as a laboratory for testing solar/stellar dynamo theory \citep[e.g.,][]{charbonneauLR,karakLRSP}.
Solar magnetic activity is dominated by quasi-periodicity with roughly an 11-year time scale \citep{hathawayLR}, called the Schwabe cycle.
The Schwabe cycle is strongly modulated by secular variability, ranging from the periods of very low solar activity, known as Grand minima \citep[the most recent being the Maunder minimum between 1645\,--\,1715 -- ][]{eddy76,usoskin_MM_15}, to periods of very high solar activity, called Grand maxima, such as in the second half of the 20th century \citep{usoskin_PRL_03,solanki_Nat_04}.
However, the statistical properties of solar Grand minima and maxima are poorly constrained by direct sunspot observations.
Several empirical rules describe solar cycle characteristics, such as Waldmeier's rule \citep{waldmeier39}, which relates the strength of a solar cycle to the length of its rising phase.
These relations were inferred from direct instrumental observations of sunspots over the last four centuries, covering only 25 reasonably resolved cycles and a dozen additional, more poorly constrained ones \citep{clette16}.
Consequently, the statistical significance of such empirical relations remains limited.
In particular, the question of whether solar cycles operate as a phase-locked clock or rather as a sequence of independent dynamo pulses has remained open \citep{weisshaar23}.

Indirect proxies in the form of cosmogenic isotopes, mostly $^{14}$C measured in tree rings and $^{10}$Be in ice cores, made it possible to reconstruct solar activity on the multi-millennial timescale over the Holocene \citep[e.g.,][]{solanki_Nat_04,vonmoos06,steinhilber12,beer12,wu18}.
However, this method, while reliably reconstructing long-term variability, could not resolve individual solar cycles due to the insufficient temporal resolution of isotope measurements \citep[e.g.,][]{usoskin_LR_23}.
For example, although several dozen Grand minima and maxima were found during the Holocene \citep{usoskin_AA_07,inceoglu15}, the behaviour of very low solar activity was known only for the Maunder minimum \citep[e.g.,][]{vaquero15}.

A major breakthrough in the quality of $^{14}$C measurements was made recently, thanks to the development of accelerator mass spectrometry, which made it possible to reconstruct individual solar cycles from annual tree-ring data \citep[see a review by][]{Heaton24}.
The first of this kind was a precision set of annual $^{14}$C measurements covering the last millennium \citep{brehm21}.
A careful analysis of this dataset allowed us to reconstruct 85 individual sunspot cycles for the period 970\,--\,1900 CE \citep{usoskin_AA_21}.
A similar precision dataset for the first millennium BCE \citep{brehm25} led to a reconstruction of 93 solar cycles for that period \citep{usoskin_AA_25}.
These results provided important constraints on solar dynamo theory \citep[e.g.,][]{biswas23}. 

Very recently, the millennium-long gap between the previously measured annual-resolution datasets has been bridged by a high-resolution $^{14}$C dataset covering the first millennium CE \citep{wang_14C_26}.
By applying the methodology developed by us earlier \citep{usoskin_AA_21,usoskin_AA_25} to this new dataset, we reconstruct individual sunspot cycles for the period 1\,--\,970 CE and present the results of their statistical analysis.

\section{Data}
\label{S:methods}

\label{Ss:data}

Several datasets have been used in this work, as described below.

\begin{figure}[t]
   \centering
   \includegraphics[width=\hsize]{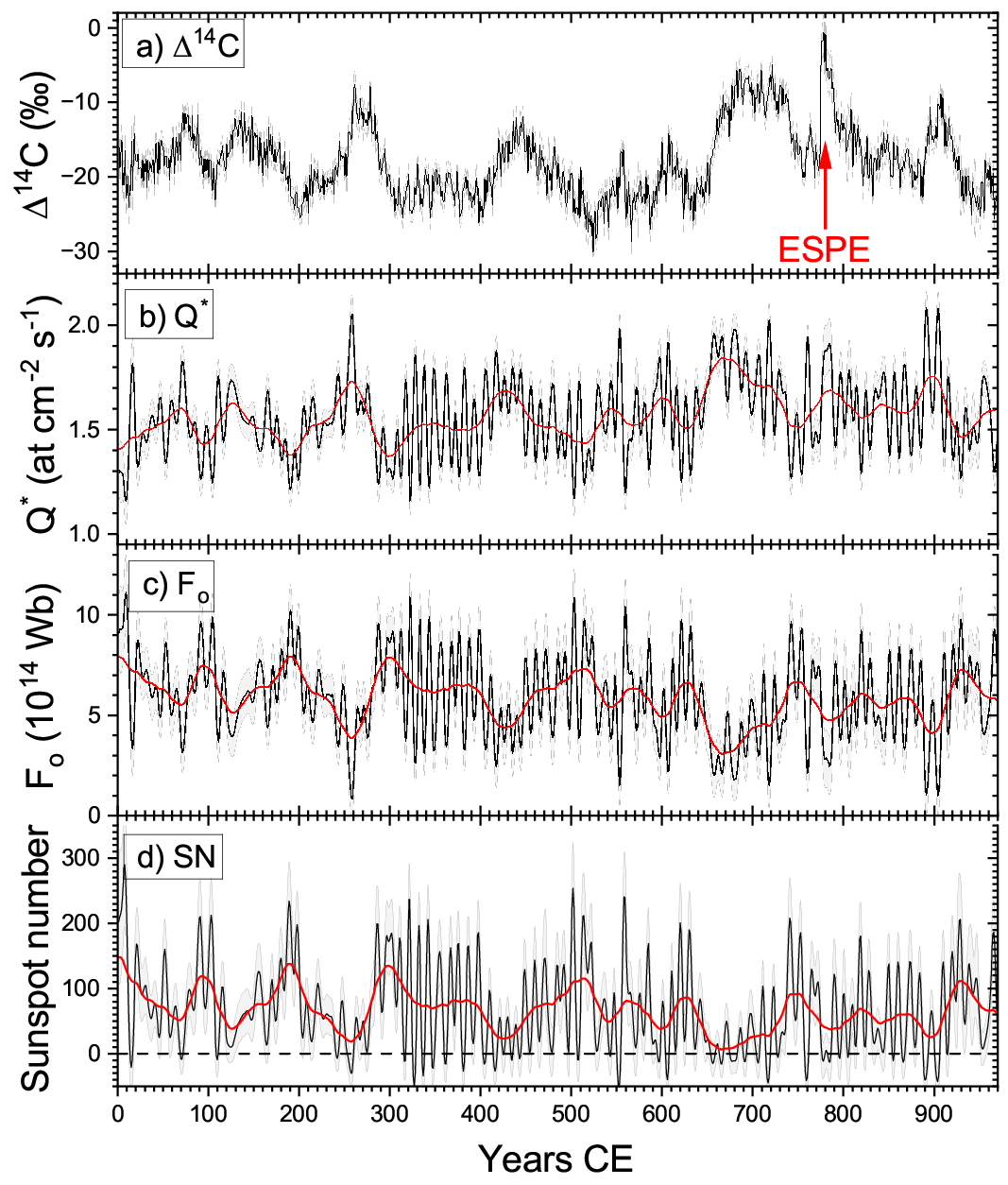}
      \caption{Sequential steps of the annual solar activity reconstructions for the first millennium CE.
      Panel a) Raw $\Delta^{14}$C data \citep{wang_14C_26}. The red arrow indicates the signature of the extreme solar particle event (ESPE) of 774 CE.
      Panel b) $^{14}$C production rate $Q^*$ corrected for the geomagnetic shielding and ESPE effects.
      Panel c) Open solar flux $F_{\rm o}$.
      Panel d) Sunspot number.
      Black curves, grey shading and red curves depict the annual values, 1$\sigma$ (68\% confidence interval) uncertainties, and 22-year smoothed data, respectively.}
         \label{fig:1000}
   \end{figure}

Our main data set is the series of the relative concentration of radiocarbon $^{14}$C, measured at an annual resolution in European oak trees for the period 1\,--\,969 CE \citep{wang_14C_26}.
It is shown in Figure~\ref{fig:1000}a along with the $1\sigma$ error bars provided, which are of the order of 1.6 \textperthousand.
For intervals with multiple measurement points, a weighted average was used to obtain annual values and their uncertainties. 
The dataset includes several short gaps (data in the years 475, 573, 804, 805, 809, 928, 929, 944, and 945 CE were missing), which were linearly interpolated, and an enhanced error bar of $\sigma$=5 \textperthousand\ was assigned to the interpolated datapoints.

\begin{figure}[t]
   \centering
   \includegraphics[width=\hsize]{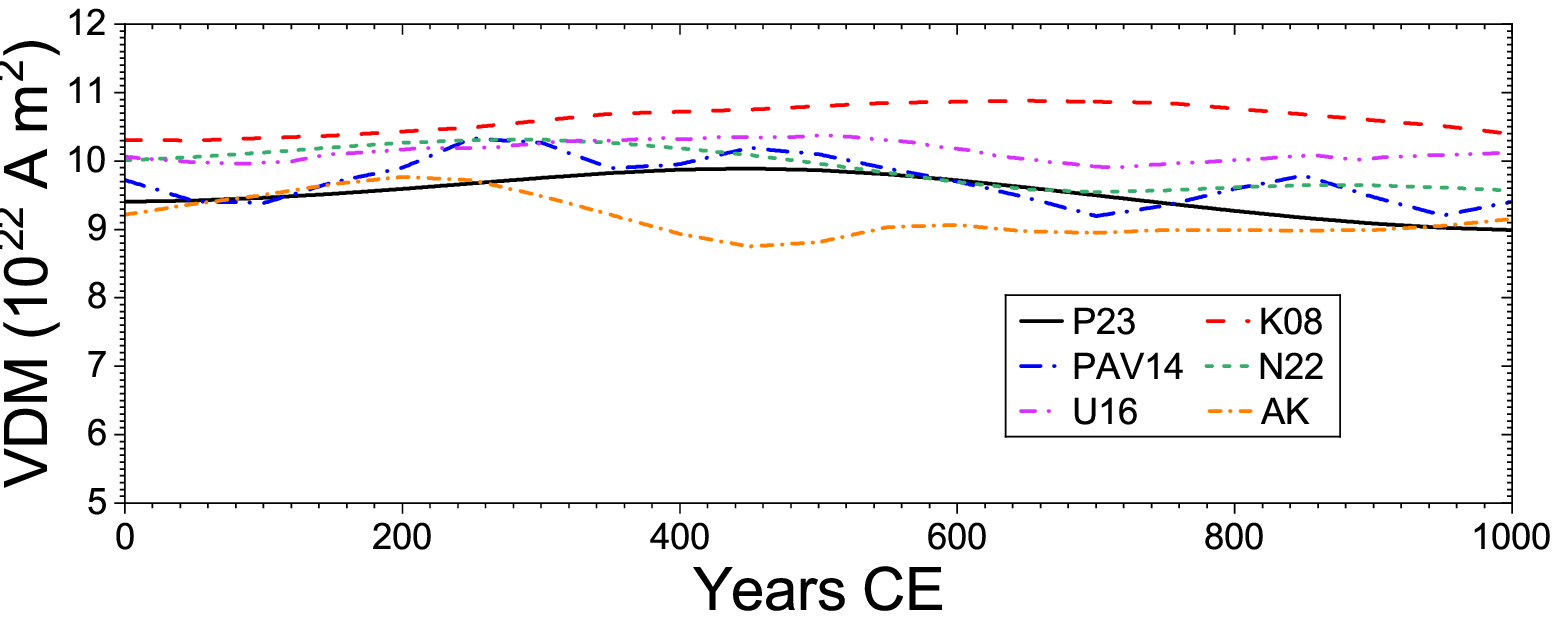}
      \caption{Recent reconstructions of the geomagnetic virtual dipole moment VDM for the first millennium CE.
      The models are: P23 \citep[][]{panovska23}; K08 \citep[][]{knudsen08}; PAV14 \citep[][]{pavon14}; N22 \citep{nilsson22}; U16 \citep{usoskin_AA_16}; AK \citep{schanner22}.}
      \label{Fig:VDM}
\end{figure}

Since the reconstruction of solar activity from cosmogenic isotopes involves correction for the geomagnetic shielding, we considered six recent archeo/paleo-magnetic models covering the period of interest, in the form of the virtual dipole moment (VDM) as shown in Figure~\ref{Fig:VDM}.
As seen, the geomagnetic field was stable and strong (VDM in the range (9\,--\,11)$\cdot$10$^{22}$ A m$^2$ bounded by the AK and K08 curves. 

\begin{figure}[t]
   \centering
   \includegraphics[width=\hsize]{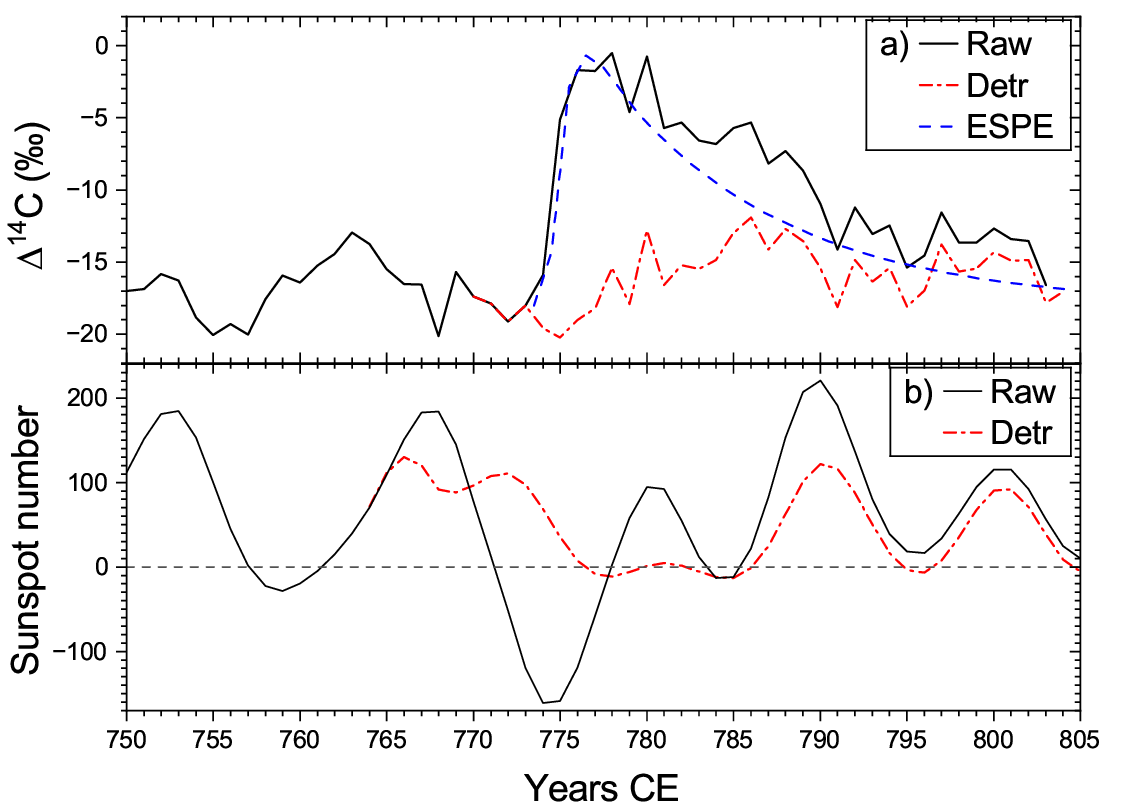}
      \caption{Correction of the dataset for the ESPE of 774 CE.
      Panel a) Mean raw $\Delta^{14}$C annual data (black curve); the modelled 774 response \citep[blue dashed lines, T=130 DoY, A=3.5, see][]{golubenko25}; detrended $\Delta^{14}$C data (red dash-dotted curve).
      Panel b) Sunspot numbers reconstructed from the raw $\Delta^{14}$C (black) and detrended (red) datasets. 
      Error bars are omitted for clarity in all panels.
      }
         \label{fig:774}
   \end{figure}

While the radiocarbon production is mostly defined by the Galactic Cosmic Rays (GCR), very roughly once per millennium the Sun produces extreme solar particle events (ESPEs) observable as distinct increases in $\Delta^{14}$C \citep[e.g.,][]{miyake12,usoskin_775_13,cliver_LR_22}.
While an ESPE leads to a nearly instantaneous production of a large amount of $^{14}$C in the atmosphere, its signature in the measured concentration $\Delta^{14}$C may spread over a decade, distorting the solar activity signal \citep[e.g.,][]{miyake19}.
If unaccounted, the enhanced $^{14}$C production is interpreted by the reconstruction method as a weak (or even unphysically negative) sunspot number, as shown in Figure~\ref{fig:774}.
Accordingly, the dataset needs to be detrended of the ESPE before further analysis.
The analysed dataset contains one known ESPE of 774 CE \citep{miyake12}, as denoted in Figure~\ref{fig:1000}a, and was detrended, as illustrated in Figure~\ref{fig:774}a.
This was done by removing the modelled ESPE signal (dashed blue line) from the raw $\Delta^{14}$C to obtain the detrended data (red dot-dashed curve).
Although the raw-data uncertainties were directly translated to the detrended dataset, the affected reconstructed solar cycles are marked as low quality.

%

\section{Methodology}
\label{S:method}

\subsection{Reconstruction}

The sunspot number (SN) reconstruction was performed following the methodology we developed for a similar task earlier \citep[see][for details]{usoskin_AA_21,usoskin_AA_25}.
The reconstruction method is based on the Markov Chain Monte Carlo (MCMC) approach with 10000 ensemble reconstructions. It is composed of four consecutive steps, each involving its own source of uncertainties: 
\begin{equation}
\Delta^{14}{\rm C} \stackrel{(1)}{\longrightarrow} Q \stackrel{(2)}{\longrightarrow} Q^* \stackrel{(3)}{\longrightarrow} F_{\rm o} \stackrel{(4)}{\longrightarrow} {\rm SN}.
\label{eq:method}
\end{equation}
Here $Q$ is the ${14}{\rm C}$ production rate, $Q^*$ is the same after correcting for the shielding by the geomagnetic field, and $F_{\rm o}$ is the open magnetic flux. 
Before processing, the data set was corrected for the effect of the ESPE of 774 CE as described above. 

In Step 1, for the $j$-th year and $i$-th ensemble realisation, the annual $\Delta^{14}$C value was taken as 
\begin{equation}
\Delta_{i,j}=\langle{\Delta}\rangle_j + R_{i,j}\cdot\sigma_{\Delta j},    
\end{equation}
where $\langle{\Delta}\rangle_j$ and $\sigma_{\Delta j}$ are the mean and the uncertainty of the radiocarbon data for the $j$-th year, and $R$ is a normally distributed random number with zero mean and unit dispersion. 
Before further analysis, the simulated data series for each realisation was smoothed with a low-pass filter (\textit{lowpass} Matlab routine with a normalised passband frequency $wc$=0.17, corresponding to about three years) to avoid amplification of noise, as discussed in detail in \citet{usoskin_AA_25}. 
Then, the $^{14}$C production rate $Q_{i,j}$ was computed from these $\Delta^{14}$C values by applying the multi-box carbon cycle model \citep{guettler15,buentgen18}.
In this way, 10 000 ensemble series of $Q_{i.j}$ were obtained.

In Step 2, the production rates $Q$, computed in Step 1, were translated to those corresponding to the current conditions (VDM=$7.75\cdot 10^{22}$ A m$^2$), further denoted as $Q^*$. 
The translation was made using the calibration curve by \citet{usoskin_AA_21}. 
For each ensemble realisation $i$, one VDM model was randomly chosen from the six considered models (see Figure~\ref{Fig:VDM}) for all years.
The evolution of $Q^*$ is shown in Figure~\ref{fig:1000}b. 
It varies between 1.2\,--\,2 atom cm$^{-2}$ s$^{-1}$ with a mean value of about 1.6 atom cm$^{-2}$ s$^{-1}$.

In Step 3, the geomagnetically corrected production rate $Q^*$ was converted into the open solar flux (OSF) $F_{\rm o}$ by applying a physics-based model \citep{usoskin_AA_21,usoskin_AA_25}, which also includes an additive uncertainty term $\sigma_F$=0.9$\cdot$10$^{14}$ Wb.
The evolution of OSF is shown in Figure~\ref{fig:1000}c. 
It ranges between (2\,--\,10)$\cdot$10$^{14}$ Wb, with a mean value of about 5.9$\times$10$^{14}$ Wb. 

In Step 4, sunspot numbers (SN) were computed from the OSF $F_{\rm o}$ by applying an empirical conversion algorithm developed by \citet{usoskin_AA_21}.
The ensemble SN reconstruction is shown in Figure~\ref{fig:1000}d. 
The dominant cyclic variability is clearly seen. 
It is analysed in the next Section.

\subsection{Uncertainties of the reconstruction}

\begin{figure}[t]
   \centering
   \includegraphics[width=\hsize]{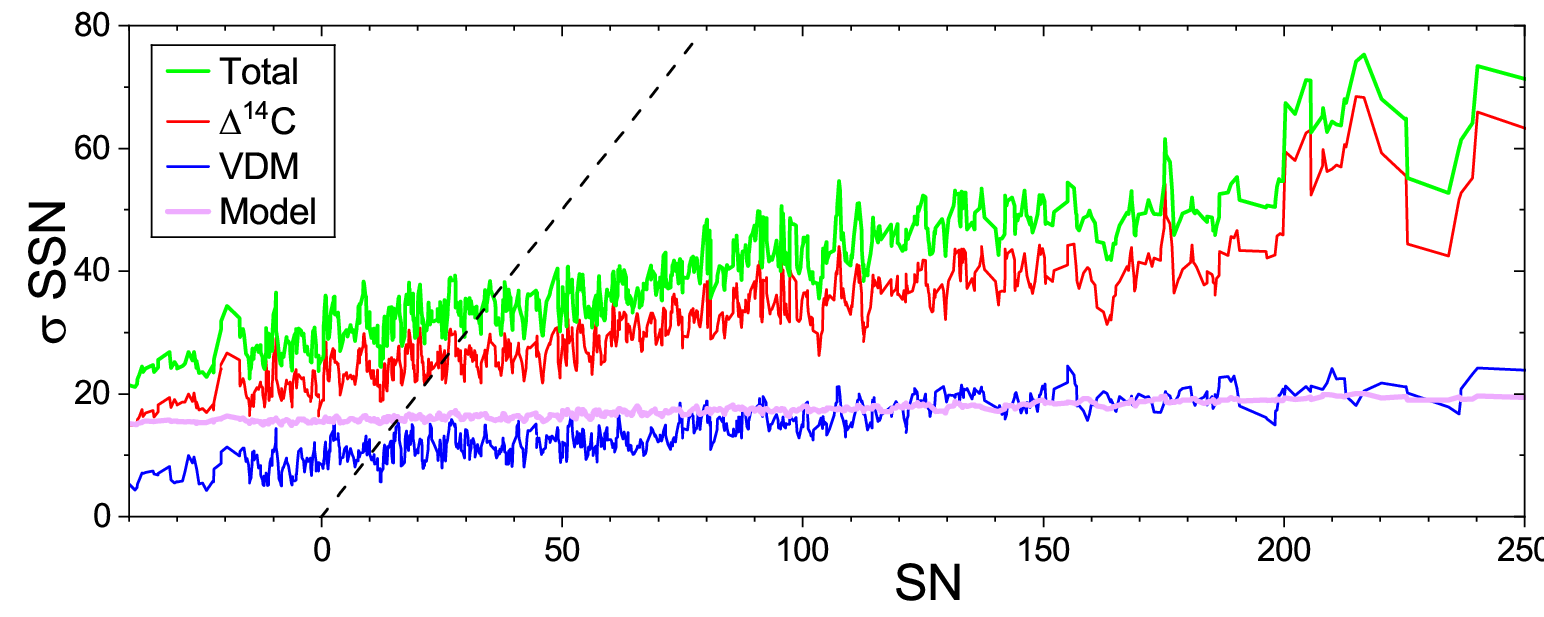}
      \caption{Different sources of uncertainties of the sunspot number reconstruction plotted at the $1\sigma$ level: radiocarbon measurement errors (red $\Delta^{14}$C line); geomagnetic models (blue VDM line); model uncertainties (magenta line) including both the computation of the OSF and OSF-to-SSN conversion; and the total uncertainty (green line), which is close to the geometrical sum of the individual ones.
      The black dashed line represents the diagonal.
      }
      \label{Fig:error}
\end{figure}

There are several sources of uncertainty in the SN reconstruction.
The MCMC method enables a direct estimation of error propagation in the reconstruction, allowing for the individual switching of different uncertainty sources.
Contribution of different processes to the summary uncertainty of the SN reconstruction is shown in Figure~\ref{Fig:error} as computed for 1000 ensemble realisations.
The total $1\sigma$ uncertainty (green line) varies between 20 and 70 in SN, depending on the solar activity level.
The total uncertainty is dominated by the radiocarbon measurement errors, which are comparable to the solar cycle amplitude in $\Delta^{14}$C, as indicated by the red line.
The uncertainties related to the paleomagnetic reconstructions are modest, being 10\,--\,20 SN, because of the fairly stable and large geomagnetic dipole strength over that period.
The uncertainty of the model, which includes both the conversion between the production rate and OSF and the OSF-to-SN conversion, remains stable at about 15 SN units \citep[cf.][]{usoskin_AA_25}.
Since the sources of uncertainty are independent of each other, the total uncertainty is very close to the geometrical sum of individual uncertainties.
As seen, the reconstructed SN is uncertain for SN$<$50, while for strong cycles, the uncertainties reach about one third of the SN value.

\section{Results and analyses}
\label{S:results}

\subsection{Reconstructed sunspot numbers}
\label{ss:SN}
\begin{figure}[t]
   \centering
   \includegraphics[width=\hsize]{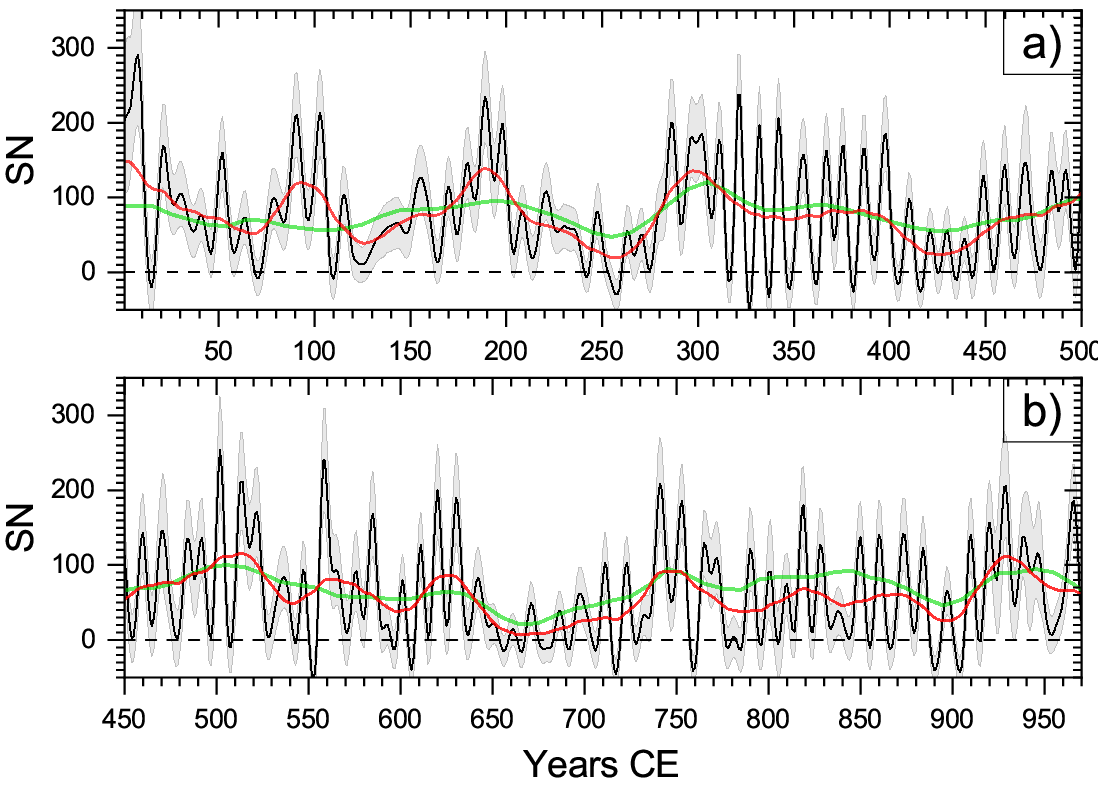}
      \caption{Sunspot numbers (ISN\_v2 normalisation)  reconstructed here for 1\,--\,969 CE, split into two panels for clarity.
      The black curve, grey shading, and red curve depict the mean annual reconstructed SN, its 68\% conﬁdence interval, and the 22-year smoothed evolution (see main text), respectively. These data are identical to those in Figure~\ref{fig:1000}d. The digital version of these data is available in CDS. The green curve depicts the smooth decadal SN values reconstructed from multi-proxy cosmogenic isotope data \citep{wu18}.
      }
      \label{Fig:SN}
\end{figure}

The reconstructed SN series, in the International Sunspot Number (ISN) v.2 definition \citep{clette16}, is shown in Figure~\ref{Fig:SN}, which is split into two panels for better visibility.
The evolution is dominated by cyclic variations in activity, which are modulated on longer timescales.
To highlight secular changes, we also show, as a red line, a smooth 22-year low-pass signal, which varies between nearly zero and 100 in SN units.
The green line depicts the decadal SN reconstruction based on a multi-proxy approach \citep{wu18}.
Unsurprisingly, the latter is smoother, being based on multi-isotope data, but shows a very similar level and long-term variability as the present reconstruction.

The analysed period contains one full Grand minimum during 650\,--\,730 CE, called the Horrebow minimum \citep{kudsk22,usoskin_LR_23}.
In addition, several shorter periods of weak activity can be observed ca. 250 CE and 430 CE.
However, no periods, similar to the modern Grand maximum, with the maximum SN systematically exceeding 200, appear in the reconstructed series.
The solar activity level was more stable (11 weak cycles with $\langle$SN$\rangle<20$) during the first millennium CE than during the first millennium BCE \citep[16 weak cycles, two Grand minima --][]{usoskin_AA_25} and second millennium CE \citep[42 weak cycles, three Grand minima --][]{usoskin_AA_21}.
This implies that Grand minima are not evenly distributed in time but tend to cluster over a multi-millennial scale \citep{usoskin_AA_07,usoskin_AA_16}.

\begin{figure}[t]
   \centering
   \includegraphics[width=1.1\hsize]{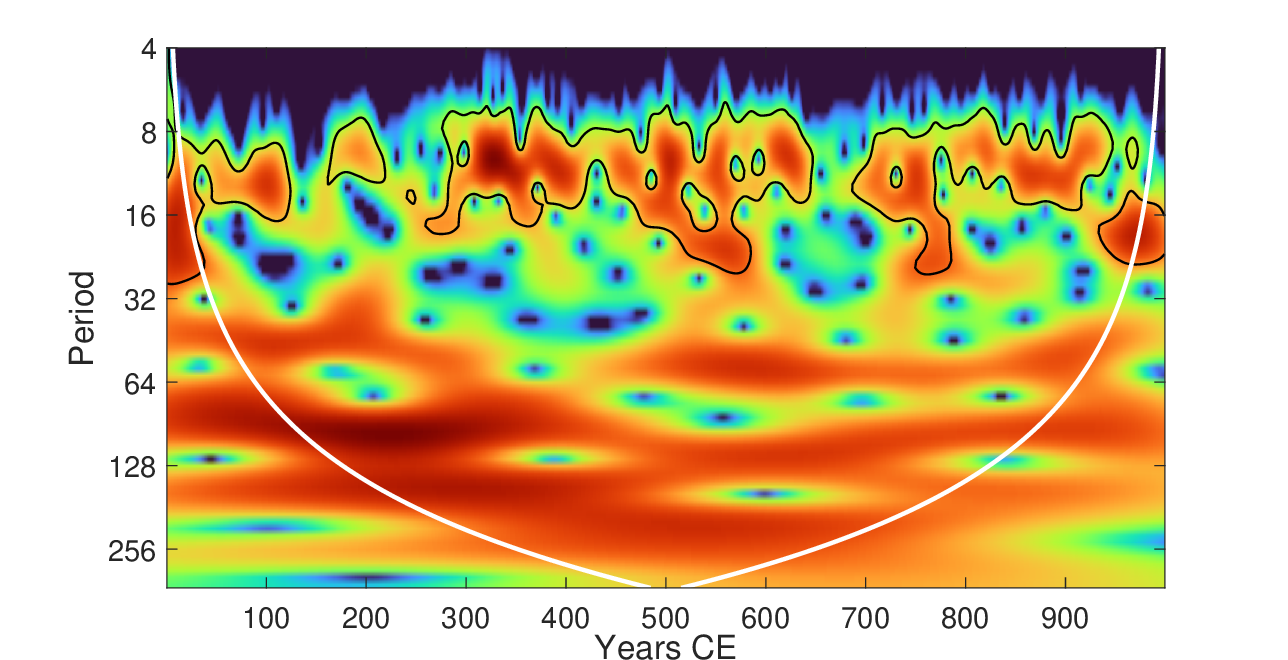}
      \caption{Wavelet spectrum (Morlet basis, $k$=6) of the reconstructed annual mean SN series. The black-contoured areas represent the statistically significant spectral power ($p<$ 0.05 against the AR1 red noise). The white line bounds the cone of influence below which the results are unreliable.}
         \label{fig:wv}
   \end{figure}

A wavelet spectrum of the reconstructed sunspot series is shown in Figure~\ref{fig:wv}.
The only significant periodicity is around 11 years, with a broad range of periods between 8 and 15 years.
However, the 11-year periodicity appears insignificant around the periods of weak activity ca. 150, 250 and 700 CE.
Other notable, but not statistically significant quasi-periodicities are found around 200 years (Suess or de Vries cycle), 70\,--\,140 years (Geissberg cycle), and 50 years.

%

\subsection{Solar cycles}

From the annually reconstructed sunspot numbers (Figure~\ref{Fig:SN}), one can identify individual solar cycles.
A total of 91 solar cycles were identified as presented in Table~\ref{tab:cycles}.
The quality of the reconstructed cycles is uneven and may vary from poorly resolved to very clear solar cycles. 
To account for that, we use, similarly to previous studies \citep{usoskin_AA_21,usoskin_AA_25}, the quality flag $q$ defined as follows:
\begin{enumerate}
\setcounter{enumi}{-1}
\item The cycle cannot be identified, but only guessed, or its strength is consistent with zero at the 1$\sigma$ level. Ten such cycles were found.
\item  The cycle is strongly distorted, or its strength is consistent with zero at the 2$\sigma$ level. The timing of cycle minima and maxima cannot be reliably defined. We found 12 such cycles.
\item  The cycle can be approximately identified, but either its shape or level is distorted. The timing of cycle minima can be imprecise for several years, while the maximum can be undefined. We found 18 such cycles.
\item  The cycle is reasonably well defined, but the minimum or maximum is not clear, or the cycle contains a gap longer than one year. We found 24 such cycles.
\item  The cycle is clear in shape, but has a somewhat unclear amplitude. We found 15 such cycles.
\item  The cycle is clear in shape and amplitude. We found 11 such cycles.
\end{enumerate}
Out of 91 identified cycles, 26 are well defined ($q>3$), 24 are reasonably defined ($q$=3), 30 are poorly defined ($q=$1\,--\,2), and 10 are undefined or only guessed ($q$=1).
The first and the last cycles could be affected by the edges of the data series.

\begin{figure}[t]
   \centering
   \includegraphics[width=\hsize]{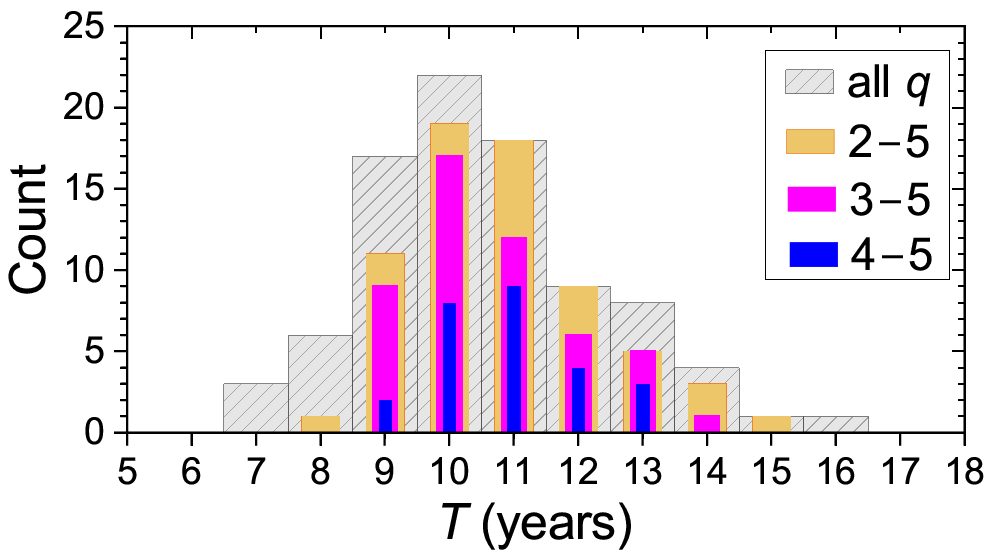}
      \caption{Histogram of the solar cycle lengths $T$ (min-to-min) for all the reconstructed cycles (grey shaded bars) and for cycles corresponding to different values of the quality flag $q$ as indicated in the legend.
      }
      \label{Fig:T}
\end{figure}

With a total of 91 cycles during 970 years, the average cycle length is 10.6 years, which is fully consistent with the cycle deduced from directly observed sunspot records and also with the cycle obtained from cosmogenic isotopes. 
E.g., \citet{usoskin_AA_21} found a mean length of 10.8 years for the second millennium CE, while \citet{usoskin_AA_25} obtained 10.5 years for the first millennium BCE.
However, individual cycles vary in length quite a bit, as shown in Figure~\ref{Fig:T}. 
The cycle length varies from 7 to 16 years if we consider all cycles, but this range is tighter for well-defined cycles, viz. 9\,--\,13 years for $q>$3, and 8\,--\,15 years for $q>$2.
This is consistent with cycles both directly observed after 1750 and reconstructed for the last millennium.

The validity of empirical relations, including the Waldmeier rule, remains inconclusive for the reconstructed cycles, likely due to uncertainties in the cycle shape.

\begin{table*}[t]
\caption{Reconstructed solar cycles and their parameters: consequent number $n$; years CE of the minimum and maximum $Y_{\rm min}$ and $Y_{\rm max}$, respectively; the cycle average sunspot number $\langle$SN$\rangle$ with its 68\% confidence intervals; Quality flag $q$; and comments. }
\small
    \centering
    \begin{tabular}{crrccc|crrccc}
    \hline
n & Y$_{\rm min}$ & Y$_{\rm max}$ & $\langle$SN$\rangle$ & $q$ & Comments & n & Y$_{\rm min}$ & Y$_{\rm max}$ & $\langle$SN$\rangle$ & $q$ & Comments \\
\hline
1 & 4 & 8 & 182$\pm$25 & 2 &  & 47 & 497 & 502 & 110$\pm$17 & 4 &  \\
2 & 15 & 21 & 92$\pm$14 & 2 &  & 48 & 508 & 514 & 127$\pm$18 & 3 &  \\
3 & 27 & 31 & 87$\pm$13 & 1 &  & 49 & 518 & 522 & 115$\pm$17 & 2 &  \\
4 & 37 & 41 & 60$\pm$12 & 1 &  & 50 & 528 & 536 & 48$\pm$10 & 2 &  \\
5 & 46 & 52 & 88$\pm$12 & 4 &  & 51 & 542 & 547 & 26$\pm$10 & 2 &  \\
6 & 58 & 64 & 42$\pm$9 & 3 &  & 52 & 553 & 558 & 116$\pm$17 & 2 &  \\
7 & 70 & 79 & 54$\pm$10 & 2 &  & 53 & 564 & 565 & 64$\pm$18 & 0 & gap 573 CE \\
8 & 84 & 91 & 133$\pm$14 & 3 &  & 54 & 571 & 576 & 53$\pm$15 & 2 &  \\
9 & 97 & 103 & 117$\pm$13 & 3 &  & 55 & 580 & 585 & 92$\pm$15 & 3 &  \\
10 & 110 & 115 & 53$\pm$11 & 3 &  & 56 & 590 & 593 & 10$\pm$11 & 0 &  \\
11 & 121 & xx & 14$\pm$9 & 0 &  & 57 & 597 & 601 & 27$\pm$11 & 3 &  \\
12 & 130 & xx & 47$\pm$11 & 0 &  & 58 & 606 & 611 & 49$\pm$13 & 4 &  \\
13 & 139 & xx & 60$\pm$11 & 0 &  & 59 & 615 & 620 & 99$\pm$16 & 5 &  \\
14 & 149 & 155 & 86$\pm$10 & 2 &  & 60 & 625 & 630 & 91$\pm$15 & 4 &  \\
15 & 164 & 170 & 67$\pm$12 & 3 &  & 61 & 637 & 642 & 47$\pm$12 & 2 &  \\
16 & 174 & 180 & 102$\pm$14 & 3 &  & 62 & 648 & 651 & 11$\pm$9 & 0 & Horrebow minimum \\
17 & 184 & 189 & 169$\pm$18 & 3 &  & 63 & 656 & 661 & 1$\pm$7 & 1 & Horrebow minimum \\
18 & 194 & 198 & 132$\pm$15 & 4 &  & 64 & 665 & 671 & 12$\pm$7 & 1 & Horrebow minimum \\
19 & 204 & 208 & 51$\pm$11 & 3 &  & 65 & 678 & 686 & 8$\pm$7 & 1 & Horrebow minimum \\
20 & 214 & 221 & 70$\pm$10 & 2 &  & 66 & 691 & 697 & 27$\pm$8 & 1 & Horrebow minimum \\
21 & 226 & xx & 58$\pm$11 & 0 &  & 67 & 705 & 711 & 36$\pm$9 & 2 & Horrebow minimum \\
22 & 234 & xx & 31$\pm$9 & 0 &  & 68 & 717 & 723 & 31$\pm$8 & 2 & Horrebow minimum \\
23 & 243 & 248 & 31$\pm$8 & 1 &  & 69 & 728 & 733 & 20$\pm$10 & 0 &  \\
24 & 256 & 263 & 11$\pm$9 & 2 &  & 70 & 736 & 741 & 126$\pm$16 & 3 &  \\
25 & 267 & 271 & 32$\pm$12 & 2 &  & 71 & 747 & 753 & 92$\pm$12 & 3 &  \\
26 & 275 & 286 & 101$\pm$12 & 1 &  & 72 & 759 & 765 & 57$\pm$12 & 1 &  \\
27 & 291 & 296 & 129$\pm$18 & 1 &  & 73 & 769 & 772 & 67$\pm$15 & 1 & ESPE 774 CE \\
28 & 299 & 302 & 155$\pm$19 & 1 &  & 74 & 778 & 781 & -4$\pm$11 & 0 & ESPE 774 CE \\
29 & 307 & 311 & 109$\pm$15 & 3 &  & 75 & 785 & 790 & 51$\pm$14 & 3 &  \\
30 & 316 & 321 & 89$\pm$13 & 5 &  & 76 & 796 & 801 & 45$\pm$18 & 4 & gap 804 CE \\
31 & 327 & 332 & 68$\pm$13 & 5 &  & 77 & 805 & 810 & 36$\pm$16 & 3 & gap 805,809 CE \\
32 & 337 & 342 & 73$\pm$13 & 5 &  & 78 & 814 & 819 & 90$\pm$15 & 5 &  \\
33 & 348 & 355 & 72$\pm$11 & 5 &  & 79 & 824 & 828 & 68$\pm$14 & 4 &  \\
34 & 361 & 367 & 74$\pm$13 & 4 &  & 80 & 834 & 838 & 24$\pm$10 & 2 &  \\
35 & 371 & 376 & 87$\pm$14 & 4 &  & 81 & 844 & 850 & 63$\pm$12 & 4 &  \\
36 & 381 & 387 & 74$\pm$12 & 5 &  & 82 & 855 & 860 & 59$\pm$12 & 5 &  \\
37 & 392 & 398 & 88$\pm$12 & 5 &  & 83 & 866 & 873 & 64$\pm$11 & 4 &  \\
38 & 404 & 410 & 33$\pm$9 & 5 &  & 84 & 879 & 884 & 49$\pm$11 & 4 &  \\
39 & 416 & 421 & 18$\pm$9 & 3 &  & 85 & 890 & 897 & 8$\pm$7 & 3 &  \\
40 & 425 & 430 & 22$\pm$10 & 3 &  & 86 & 904 & 910 & 47$\pm$12 & 4 &  \\
41 & 435 & 439 & 16$\pm$10 & 3 &  & 87 & 915 & 920 & 91$\pm$17 & 3 & gap 928 CE \\
42 & 444 & 449 & 66$\pm$13 & 4 &  & 88 & 924 & 929 & 133$\pm$26 & 3 & gap 929 CE \\
43 & 454 & 460 & 69$\pm$14 & 5 &  & 89 & 934 & 939 & 102$\pm$22 & 2 & gap 944,945 CE \\
44 & 465 & 471 & 78$\pm$16 & 4 & gap 475 CE & 90 & 943 & 946 & 72$\pm$16 & 2 &  \\
45 & 478 & 484 & 73$\pm$15 & 3 &  & 91 & 954 & 965 & 77$\pm$9 & 1 &  \\
46 & 488 & 492 & 88$\pm$17 & 3 &  &  &  &  &    &  &  \\
\hline
    \end{tabular}
    \label{tab:cycles}
\end{table*}

\section{Conclusions}
\label{s:conclusions}

An annual reconstruction of solar activity in the form of the sunspot number (in the ISN v.2 definition) is presented for the first millennium CE, viz. 1\,--\,969 CE.
The reconstruction is based on the annual measurements of $\Delta^{14}$C by \citet{wang_14C_26} and involves an MCMC reconstruction method \citep{usoskin_AA_21,usoskin_AA_25}, and accounts for different sources of uncertainties in a straightforward manner. 
The reconstructed series covers one Grand solar minimum, called the Horrebow minimum, during 650\,--\,730 CE, which is comparable to the Maunder minimum, and a few shorter episodes of low activity occurring ca. 120, 250, and 900 CE, resembling the Dalton minimum. 
The time-frequency analysis, using the wavelet spectrum, shows that the only significant quasi-periodicity in the reconstructed series is related to the Schwabe cycle. 
Other periodicities, such as the Suess/de Vries cycles, Gleissberg variability or its harmonics, appear insignificant.
There was a strong ESPE in 774 CE, the strongest one known during the Holocene.
The data were corrected for it, but reconstructed cycles 73\,--\,75 may well still be affected by the event.

Based on this SN reconstruction, we identified 91 solar cycles during that period, of which 26 are well defined ($q\geq 4$), 24 are reasonably defined ($q$=3), 31 poorly defined ($q=$1\,--\,2), and 10 can only be guessed.
This dataset provides no conclusive evidence for the existence of Waldmeier's relations. 

The results of this work fill the gap in the solar cycle statistics between the first millennium BCE \citep{brehm21,usoskin_AA_21} and the second millennium CE \citep{brehm25,usoskin_AA_25}.
The total number of resolved solar cycles now is about 280, which is much greater than the 25 directly observed cycles in the ISN data series \citep{clette16}.
This provides greatly improved statistical constraints on solar dynamo theory as well as on the models of solar irradiance on the millennial timescale.
However, two caveats should be considered: 
1) Many of the reconstructed cycles are rather uncertain. If we consider only the reliably reconstructed cycles, i.e. those with $q\geq 4$, then we have a smaller number of 74 cycles, which, however, is still three times larger than the directly measured ones. 
2) This collective statistic is based on three stand-alone pieces of millennium-long reconstructions with possible boundary distortions between them.
A consistent 3-millennium-long reconstruction of sunspot numbers, based on a composite multi-proxy cosmogenic record, is pending.

\begin{acknowledgements}
This work has received funding from the European Research Council for Synergy Grant (project 101166910) and Advanced Grant (grant agreement No. 101097844 — project WINSUN). 
\end{acknowledgements}

%

\bibliographystyle{aa}
\bibliography{usoskin_all}

\end{document}